\documentclass[12pt]{article}
\usepackage{amsmath}
\def\be{\begin{equation}}
\def\ee{\end{equation}}
\def\arr{\begin{array}{rll}}
\def\ea{\end{array}}
\def\bea{\begin{eqnarray}}
\def\eea{\end{eqnarray}}

\def\N2{$N{=}2$}

\def\>{\rangle}
\def\<{\langle}
\def\+{\dagger}
\def\={\ =\ }

\begin{document}
\renewcommand{\thefootnote}{\fnsymbol{footnote}}
\begin{titlepage}
\setcounter{page}{0}
\begin{flushright}
LMP-TPU--7/11  \\
\end{flushright}
\vskip 1cm
\begin{center}
{\LARGE\bf Remarks on $l$--conformal  extension}\\
\vskip 0.5cm
{\LARGE\bf of the Newton-Hooke algebra}\\
\vskip 1cm
$
\textrm{\Large Anton Galajinsky and Ivan Masterov\ }
$
\vskip 0.7cm
{\it
Laboratory of Mathematical Physics, Tomsk Polytechnic University, \\
634050 Tomsk, Lenin Ave. 30, Russian Federation} \\
{E-mails: galajin@mph.phtd.tpu.ru, masterov@mph.phtd.tpu.ru}

\end{center}
\vskip 1cm
\begin{abstract} \noindent
The $l$--conformal extension of the Newton--Hooke algebra proposed in [J. Math. Phys. 38 (1997) 3810]
is formulated in the basis in which the flat space limit is unambiguous.
Admissible central charges are specified. The infinite--dimensional Virasoro--Kac--Moody type extension
is given.
\end{abstract}

\vskip 1cm
\noindent
PACS numbers: 11.30.-j, 11.25.Hf, 02.20.Sv

\vskip 0.5cm

\noindent
Keywords: conformal Galilei algebra, conformal Newton--Hooke algebra

\end{titlepage}

\renewcommand{\thefootnote}{\arabic{footnote}}
\setcounter{footnote}0

Current studies of the non--relativistic version of the AdS/CFT correspondence stimulate a renewed interest in non--relativistic conformal
algebras. In general, one can formulate a family of conformal Galilei algebras which are parameterized by a positive half integer $l$ \cite{nor,nor1}
(see also \cite{hen}). Furthermore, following the idea in \cite{hen1}, one can extend
the $l$--conformal Galilei algebra\footnote{In modern literature the reciprocal of $l$ is called the rational dynamical exponent.
The corresponding
algebra is referred to as
the conformal Galilei algebra with rational dynamical exponent (see e.g. \cite{dh1}). In this work we use the terminology
originally adopted in \cite{nor,nor1}.} to the infinite--dimensional Virasoro--Kac--Moody type algebra \cite{bg,mt}. So far, the instances of
$l=1/2$ (the Schr\"odinger algebra) and $l=1$ (the conformal Galilei algebra) have been the focus of most studies.

In the chain of the $l$--conformal Galilei algebras $l=1$ is the only instance where the temporal and spatial coordinates scale the same way
under dilatations. This case has been extensively investigated recently in regard to dynamical realizations \cite{lsz,fil}, the
AdS/CFT correspondence \cite{bg,mt,adv}, and the Newton--Cartan structures \cite{dh1,dh}. For a review and further references
see \cite{dh1}.

According to the analysis in \cite{dh1}, a proper arena where the conformal Galilei groups act is the Newton-Hooke spacetime with
quantized negative cosmological constant. This proposal calls for a better understanding of the $l$--conformal Newton--Hooke algebra.

The $l$--conformal Newton--Hooke algebra has been previously studied in \cite{nor,nor1}. However, the flat space limit is problematic in the formulation of \cite{nor,nor1}.
This is to be contrasted with the Newton--Hooke extension of the ordinary Galilei algebra (see e.g. \cite{bac,gp}) where the latter follows from
the former in the limit in which the cosmological constant tends to zero.

The purpose of this brief note is to formulate the $l$--conformal Newton--Hooke algebra
in the basis in which the flat space limit is unambiguous. Our strategy is to generalize the analysis of $l=1/2$ \cite{gala,gala1} to the
case of arbitrary positive half integer $l$. We also analyze admissible central charges and construct
the infinite--dimensional Virasoro--Kac--Moody type extension.

Consider first a realization of the $l$--conformal Galilei algebra
\begin{align}\label{algebra}
&
[H,D]=H, &&  [H,C^{(n)}_i]=n C^{(n-1)}_i,
\nonumber\\[2pt]
&
[H,K]=2 D, && [D,K]=K,
\nonumber\\[2pt]
&
[D,C^{(n)}_i]=(n-l) C^{(n)}_i, && [K,C^{(n)}_i]=(n-2l) C^{(n+1)}_i,
\nonumber\\[2pt]
&
[M_{ij},C^{(n)}_k]=-\delta_{ik} C^{(n)}_j+\delta_{jk} C^{(n)}_i, && [M_{ij},M_{kl}]=-\delta_{ik} M_{jl}-\delta_{jl} M_{ik}+
\delta_{il} M_{jk}+\delta_{jk} M_{il},
\end{align}
in a flat $(d+1)$--dimensional
spacetime \cite{nor1}
\bea\label{gen}
&&
H=\partial_t, \qquad \qquad D=t \partial_t+l x_i \partial_i, \qquad \qquad K=t^2 \partial_t+2 l t x_i \partial_i,
\nonumber\\[2pt]
&&
C^{(n)}_i=t^n \partial_i, \qquad M_{ij}=x_i \partial_j-x_j \partial_i.
\eea
Here $n=0,1,\dots, 2l$ and  $i=1,\dots,d$.
As usual, $\partial_t=\frac{\partial}{\partial t}$, $\partial_i=\frac{\partial}{\partial x_i}$ and
summation over repeated indices is understood.
The operators $H$, $D$ and $K$ generate time translations, dilatations and special
conformal transformations, respectively. The instances of $n=0$ and $n=1$ in $C^{(n)}_i$ produce space translations and the Galilei boosts.
Higher values of $n$ correspond to accelerations. Below we first treat the case of a negative cosmological
constant and then consider a positive cosmological constant.

In order to construct the Newton--Hooke extension of (\ref{algebra}), we introduce a dimensionful constant $R$,
the characteristic time\footnote{The characteristic time is related to the cosmological constant via $\Lambda\sim\pm \frac{1}{(c R)^2}$ where $c$ is the speed of light.} \cite{bac,gp},
and deform $D$ and $K$ by analogy with the $l=1/2$ case considered in \cite{gala}
\bea\label{gener}
&&
D=\frac 12 R \sin{(2t/R)} \partial_t+l \cos{(2t/R)} x_i \partial_i,
\nonumber\\[2pt]
&&
K=-\frac 12 R^2 (\cos{(2t/R)}-1) \partial_t+l R \sin{(2t/R)} x_i \partial_i.
\eea
The generators of time translations and space rotations maintain their form while $C^{(n)}_i$ is modified as follows
\be\label{gener1}
C^{(n)}_i=R^n {(\tan{(t/R)})}^n {(\cos{(t/R)})}^{2l} \partial_i.
\ee
It is readily verified that these operators form a closed algebra. As compared to (\ref{algebra}), only the first line is modified to
include contributions which involve the cosmological constant
\begin{align}\label{algebra1}
&
[H,D]=H-\frac{2}{R^2} K, && [H,C^{(n)}_i]=n C^{(n-1)}_i+\frac{(n-2l)}{R^2} C^{(n+1)}_i.
\end{align}
Thus, for the case of a negative cosmological constant the structure relations of the $l$--conformal Newton--Hooke algebra
are given by (\ref{algebra}) with the first line modified in accord with (\ref{algebra1}).

The situation looks similarly in spacetime with universal cosmological repulsion. First one defines $D$, $K$ and $C^{(n)}_i$
\bea\label{gener2}
&&
D=\frac 12 R \sinh{(2t/R)} \partial_t+l \cosh{(2t/R)} x_i \partial_i,
\nonumber\\[2pt]
&&
K=\frac 12 R^2 (\cosh{(2t/R)}-1) \partial_t+l R \sinh{(2t/R)} x_i \partial_i
\nonumber\\[2pt]
&&
C^{(n)}_i=R^n {(\tanh{(t/R)})}^n {(\cosh{(t/R)})}^{2l} \partial_i,
\eea
and then determines the structure relations of the corresponding Newton--Hooke algebra.
As compared to (\ref{algebra}), only the first line gets altered
\begin{align}\label{algebra2}
&
[H,D]=H+\frac{2}{R^2} K, && [H,C^{(n)}_i]=n C^{(n-1)}_i-\frac{(n-2l)}{R^2} C^{(n+1)}_i.
\end{align}
These relations specify the $l$--conformal Newton--Hooke algebra for the case of a positive cosmological constant.

A few comments are in order. First, in the limit of a vanishing cosmological constant (i.e. $R \to \infty$)
the Newton--Hooke extensions reduce to the $l$--conformal Galilei algebra. The representations in terms of the differential
operators merge as well. This is to be contrasted with the basis chosen in \cite{nor,nor1} where the flat space limit
is problematic.

Second, viewed as formal Lie algebras, the $l$--conformal Galilei algebra and its Newton--Hooke counterpart are isomorphic.
The linear change of the basis
\be\label{H}
H \quad \rightarrow \quad H\mp \frac{1}{R^2} K
\ee
where the upper/lower sign corresponds to a negative/positive cosmological constant, yields the $l$--conformal Galilei algebra.
In this sense  Eqs. (\ref{gen}) and (\ref{gener}), (\ref{gener1}), (\ref{gener2}), (\ref{H}) can be viewed as providing
representations of the $l$--conformal Galilei algebra in flat space and in the Newton--Hooke spacetime, respectively.
It should be remembered, however, that, as far as dynamical realizations are concerned, (\ref{H}) is a change of the Hamiltonian
which alters the dynamics. In other words, a relation between two dynamical systems holds only locally \cite{nied1}
(in this respect see also \cite{nor1,dgh,zh}).\footnote{For a related discussion in the context of conformal many--body quantum mechanics see \cite{gm}.}
Comparing Eqs. (\ref{gen}) and (\ref{gener}), (\ref{gener1}), (\ref{H}), one can readily construct a coordinate transformation
(the prime denotes coordinates parameterizing flat space)
\be\label{nied}
t'=R \tan{(t/R)}, \qquad x'_i={(\cos{(t/R)})}^{-2l} x_i,
\ee
which brings (\ref{gen}) to (\ref{gener}), (\ref{gener1}), (\ref{H}) (see also a related discussion in \cite{dh1}). The instance of $l=1/2$ reproduces Niederer's oscillator coordinates \cite{nied1}.
Similar formulae hold also for the case of a positive cosmological constant.
As a byproduct, dynamical realizations of the $l$--conformal Newton--Hooke algebra can be derived from those of the
$l$--conformal Galilei algebra by redefining the Hamiltonian $H \rightarrow H\pm \frac{1}{R^2} K$, the latter term
typically providing the harmonic potential.

Third, the $l$--conformal Galilei algebra admits the infinite--dimensional
Virasoro--Kac--Moody type extension \cite{bg,mt} (for earlier studies see \cite{hen1,hp}). It can be realized in terms of
the operators\footnote{Our conventions are slightly different
form those in \cite{bg,mt}.}
\be
L^{(n)}=t^{n+1} \partial_t+l (n+1) t^n x_i \partial_i, \quad C^{(n)}_i=t^n \partial_i, \quad M^{(n)}_{ij}=t^n (x_i \partial_j-x_j \partial_i),
\ee
where $n$ is an arbitrary integer. The structure relations read
\bea\label{algebra3}
&&
[L^{(n)},L^{(m)}]=(m-n) L^{(n+m)}, \qquad  [L^{(n)},C^{(m)}_i]=[m-l(n+1)] C^{(n+m)}_i,
\nonumber\\[2pt]
&&
[L^{(n)},M^{(m)}_{ij}]=m M^{(n+m)}_{ij}, \qquad  \qquad [M^{(n)}_{ij},C^{(m)}_k]=-\delta_{ik} C^{(n+m)}_j+\delta_{jk} C^{(n+m)}_i
\nonumber\\[2pt]
&&
[M^{(n)}_{ij},M^{(m)}_{kl}]=-\delta_{ik} M^{(n+m)}_{jl}-\delta_{jl} M^{(n+m)}_{ik}+\delta_{il} M^{(n+m)}_{jk}+\delta_{jk} M^{(n+m)}_{il}.
\eea
In particular, $L^{(-1)}$, $L^{(0)}$, $L^{(1)}$ reproduce $H$, $D$, $K$ in (\ref{gen}). A representation of this algebra in the
Newton--Hooke spacetime can be constructed by analogy with the finite--dimensional subalgebra considered above. For a negative cosmological constant
one finds
\bea
&&
L^{(n)}=R^n {(\tan{(t/R)})}^n \Big(\frac 12 R \sin{(2t/R)} \partial_t+l [n+\cos{(2t/R)}] x_i \partial_i \Big),
\nonumber\\[2pt]
&&
C^{(n)}_i=R^n {(\tan{(t/R)})}^n {(\cos{(t/R)})}^{2l} \partial_i, \qquad   M^{(n)}_{ij}=R^n {(\tan{(t/R)})}^n (x_i \partial_j-x_j \partial_i),
\eea
while the change of the trigonometric functions by the hyperbolic ones yields a representation in spacetime with a positive cosmological constant.

Above we ignored the issue of central charges. Let us discuss them for the finite--dimensional algebra.
In arbitrary dimension the commutator of two vector generators can be modified to include the
central element
\be\label{cc}
[C^{(n)}_i,C^{(m)}_j]= \alpha(n,m) \delta_{ij},
\ee
where the constants $\alpha(n,m)=-\alpha(m,n)$ are subject to the constraints which follow from the Jacobi identities
\be\label{recur}
(n+m-2l) \alpha(n,m)=0, \qquad m \alpha(n,m-1)+n \alpha (n-1,m)=0.
\ee
The leftmost restriction implies that $\alpha(n,m)$ vanishes unless $n+m=2l$. The rightmost constraint gives linear relations
each of which intertwines two neighbors  in the chain of $\alpha(n,m)$--coefficients with $n+m=2l$.
For integer $l$ the chain involves $\alpha(l,l)=0$ which causes
all $\alpha(n,m)$ to vanish. For half integer $l$ the recurrence relation in (\ref{recur}) yields
\be
\alpha(n,m)={(-1)}^n \frac{n! m!}{(2l)!} \alpha(0,2l),
\ee
with $\alpha(0,2l)$ being arbitrary. Thus, in arbitrary dimension there is one central charge in the algebra.

In $(2+1)$--dimensions an extra contribution to the right hand side of (\ref{cc}) is admissible
\be
[C^{(n)}_i,C^{(m)}_j]= \alpha(n,m) \delta_{ij}+\beta(n,m) \epsilon_{ij},
\ee
where $\beta(n,m)=\beta(m,n)$ and $\epsilon_{ij}$ is the Levi-Civit\'a symbol. The Jacobi identities
yield restrictions which read as in (\ref{recur}) both for $\alpha(n,m)$ and $\beta(n,m)$.
In particular, $\alpha(n,m)$ and $\beta(n,m)$ vanish when $n+m\ne 2l$.
For half integer $l$ one can choose $n=m=\frac{(2l+1)}{2}$ which, in view of (\ref{recur}),
gives $\beta(\frac{(2l+1)}{2},\frac{(2l-1)}{2})=0$. Because all members in the $\beta(n,m)$--chain are linearly related to each other,
this makes them all vanish. At the same time, according to our analysis above, the coefficients $\alpha(n,m)$ are active in this case.
For integer $l$ the recurrence relation for $\beta(n,m)$ yields
 \be
\beta(n,m)={(-1)}^n \frac{n! m!}{(2l)!} \beta(0,2l),
\ee
while $\alpha(n,m)$ vanish.
Thus, in $(2+1)$--dimensions a central extension is feasible for any value of $l$, $\alpha(n,m)$ and $\beta(n,m)$ being responsible for
half integer and integer $l$, respectively.

As is well known, the Newton--Hooke extension of the ordinary Galilei algebra in $(3+1)$--dimensions can be derived from the
anti de Sitter algebra $so(2,3)$ (the de Sitter algebra $so(1,4)$) by applying the
non--relativistic limit \cite{bac}.
It is interesting to see if a similar relation can be revealed for the conformal counterpart of the Newton--Hooke algebra.
We conclude this work by discussing In\"on\"u--Wigner contraction of $so(2,4)$
which leads to the $l=1$ conformal Newton--Hooke algebra for the case of a negative cosmological constant.

Consider the structure relations of $so(2,4)$
\be
[M_{AB},M_{CD}]=\eta_{AC} M_{BD}+\eta_{BD} M_{AC}-\eta_{AD} M_{BC}-\eta_{BC} M_{AD},
\ee
where $A=0,\dots,5$ and $\eta_{AB}=\mbox{diag}(-,+,+,+,+,-)$. Let us treat the values $0$, $4$, $5$ and $i=1,2,3$ separately, and
change the basis
\bea
&&
H=\frac{2}{R} M_{05}, \qquad D=M_{04}, \qquad K=R (M_{54}+M_{05}), \qquad \tilde M_{ij}=-M_{ij},
\nonumber\\[2pt]
&&
C^{(0)}_i=\frac{1}{c R} (M_{i0}+M_{i4}), \qquad ~ C^{(1)}_i=\frac{1}{c} M_{i5}, \qquad ~ C^{(2)}_i=\frac{R}{c} (M_{i4}-M_{i0}),
\eea
where $c$ is the speed of light and $R$ is the characteristic time. It is straightforward to
compute brackets
among the new generators and verify that in the limit $c \to \infty$ they yield the structure relations of the
$l=1$ conformal Newton--Hooke algebra.
A similar contraction of $so(2,4)$ to the $l=1$ conformal Galilei algebra was discussed in \cite{lsz1}.
The case of a positive cosmological constant can be treated likewise.

To summarize, in this work the $l$--conformal Newton--Hooke algebra was formulated in the basis in which the flat space limit is unambiguous.
Admissible central charges in the algebra were determined. The infinite--dimensional Virasoro--Kac--Moody type extension was given.

It would be interesting to generalize the analysis of the central charges to the case of the infinite--dimensional algebra.
Supersymmetric extensions as well as their dynamical realizations are also worth studying.

\vspace{0.5cm}

\noindent{\bf Acknowledgements}\\

\noindent
We thank Peter Horv\'athy for useful comments. This work was supported by
the Dynasty Foundation, RFBR grant 11-02-90445, and RF Federal Program "Kadry" under the contracts
16.740.11.0469 and P2596.


\begin{thebibliography}{nn}
\bibitem{nor}
J. Negro, M.A. del Olmo, A. Rodriguez-Marco, J. Math. Phys. {\bf 38} (1997) 3786.
\bibitem{nor1}
J. Negro, M.A. del Olmo, A. Rodriguez-Marco, J. Math. Phys. {\bf 38} (1997) 3810.
\bibitem{hen}
M. Henkel, Phys. Rev. Lett. {\bf 78} (1997) 1940.
\bibitem{hen1}
M. Henkel, J. Stat. Phys. {\bf 75} (1994) 1023, hep-th/9310081.
\bibitem{dh1}
C. Duval, P. Horv\'athy, {\it
Conformal Galilei groups, Veronese curves, and Newton-Hooke spacetimes},
arXiv:1104.1502.
\bibitem{bg}
A. Bagchi, R. Gopakumar, JHEP {\bf 0907} (2009) 037, arXiv:0902.1385.
\bibitem{mt}
D. Martelli, Y. Tachikawa, JHEP {\bf 1005} (2010) 091, arXiv:0903.5184.
\bibitem{lsz}
J. Lukierski, P.C. Stichel, W.J. Zakrzewski, Phys. Lett. B {\bf 650} (2007) 203, hep-th/0702179.
\bibitem{fil}
S. Fedoruk, E. Ivanov, J. Lukierski, Phys. Rev. D {\bf 83} (2011) 085013,
arXiv:1101.1658.
\bibitem{adv}
M. Alishahiha, A. Davody, A. Vahedi, JHEP {\bf 0908} (2009) 022, arXiv:0903.3953.
\bibitem{dh}
C. Duval, P.A. Horv\'athy, J. Phys. A {\bf 42} (2009) 465206, arXiv:0904.0531.
\bibitem{bac}
H. Bacry, J.M. L\'evy--Leblond, J. Math. Phys. {\bf 9} (1968) 1605.
\bibitem{gp}
G.W. Gibbons, C.E. Patricot, Class. Quant. Grav. {\bf 20} (2003) 5225, hep-th/0308200.
\bibitem{gala}
A. Galajinsky, Nucl. Phys. B {\bf 832} (2010) 586, arXiv:1002.2290.
\bibitem{gala1}
A. Galajinsky, Phys. Lett. B {\bf 680} (2009) 510, arXiv:0906.5509.
\bibitem{nied1}
U. Niederer, Helv. Phys. Acta {\bf 46} (1973) 191.
\bibitem{dgh}
C. Duval, G. Gibbons, P. Horv\'athy, Phys. Rev. D {\bf 43} (1991) 3907, hep-th/0512188.
\bibitem{zh}
P-M. Zhang, P.A. Horv\'athy, {\it  Kohn's theorem and Galilean symmetry},
arXiv:1105.4401.
\bibitem{gm}
A. Galajinsky, I. Masterov, Phys. Lett. B {\bf 675} (2009) 116, arXiv:0902.2910.
\bibitem{hp}
P. Havas, J. Plebanski, J. Math. Phys. {\bf 19} (1978) 482.
\bibitem{lsz1}
J. Lukierski, P.C. Stichel, W.J. Zakrzewski, Phys. Lett. A {\bf 357} (2006) 1, hep-th/0511259.



\end{thebibliography}
\end{document}